\documentclass[twocolumn,twoside,slac]{revtex4}
\usepackage{graphicx}
\usepackage{fancyhdr}
\usepackage{rotating}
\usepackage{epsfig}
\begin{document}

\title{CMS Data Analysis: Current Status and Future Strategy}


\author{Vincenzo Innocente}
\affiliation{CERN, Geneva, Switzerland}

\begin{abstract}
We present the current status of CMS data analysis architecture and
describe work on future Grid-based distributed analysis
prototypes. CMS has two main software frameworks related to data
analysis: COBRA, the main framework, and IGUANA, the interactive
visualisation framework. Software using these frameworks is used today
in the world-wide production and analysis of CMS data. We describe
their overall design and present examples of their current use with
emphasis on interactive analysis. CMS is currently developing remote
analysis prototypes, including one based on Clarens, a Grid-enabled
client-server tool. Use of the prototypes by CMS physicists will guide
us in forming a Grid-enriched analysis strategy. 
The status of this work is presented, as is an outline of how we plan to leverage the power 
of our existing frameworks in the migration of CMS software to the Grid. 

\end{abstract}

\maketitle

\thispagestyle{fancy}


\section{Introduction}

Requirements on CMS software and computing resources~\cite{cmsctp} will far
exceed those of any existing high energy physics experiment, not only because
of the complexity of the detector and of the physics task but also for the size
and the distributed nature of the collaboration (today encompassing 2000 physicists
from 150 institutions in more than 30 countries) and the long time scale (20
or more years). It was widely recognised from the outset of planning for the
LHC Experiments in the mid-1990s, that the computing systems required to collect,
analyse and store the physics data would need to be distributed and global in
scope. In particular, studies of the computing industry and its expected development
showed that utilising computing resources external to CERN, at the collaborating
institutes (as had been done on a limited scale for the LEP experiments) would
continue to be an essential strategy, and that a global computing system architecture
would need to be developed.

Therefore, since 1995 the CMS computing group has been engaged in an extensive
R\&D program to evaluate and prototype a computing architecture that will allow
to perform the task of collecting, reconstructing, distributing and storing
the physics data correctly and efficiently in the LHC demanding environment,
and, at the same time, to present the user, either local or remote, with a simple
logical view of all objects needed to perform physics analysis or detector studies.

More recently, CMS undertook a major requirements and consensus building effort
to modernise this vision of a distributed computing model to a Grid-based computing
infrastructure. Accordingly, the current vision sees CMS computing as an activity
that is performed on the {}``CMS Data Grid System{}'' whose properties have
been described in considerable detail \cite{cmsgridreqs}.
The CMS Data Grid System specifies a division of labour between the various Grid projects and
the CMS core computing project.

% \begin{figure}
% \end{figure}

% Surround figure environment with turnpage environment for landscape
% figure
% \begin{figure}
% \end{figure}

\section{CMS Analysis Environment}

\begin{figure}[htp]
  \centering
  \vskip 1cm
  \epsfig{file=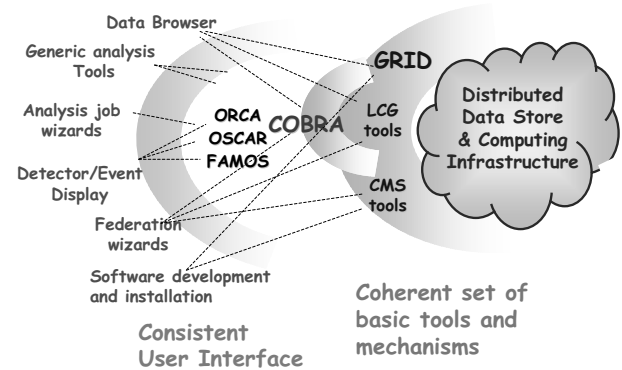,angle=-90,width=80mm} 
  \caption{CMS Analysis Environment} \label{fig01}
\end{figure}

Figure~\ref{fig01} shows the current model of the 
CMS  Analysis Environment.
 
CMS goal is to deploy a Coherent Analysis Environment
that goes beyond the interactive analysis tool
that provides classical data analysis and presentation functionalities such as
N-tuples, histograms, fitting, plotting.
We aim to allow an easy and coherent access to 
the great range of other activities typical
of the development of physics analysis and detector studies in HEP.
Support will be provided for both batch and interactive work with interface ranging
from ``pointy-clicky'' to Emacs-like power tool to scripting.
The environment will encompass
configuration management tools, application frameworks,
simulation, reconstruction and analysis packages.

Most of the activity will be seen as operations on
data-stores: Replicating entire data-stores; Copying runs,
events, event parts between stores.
A ``copy'' operation will often encompass 
something more complicated such as filtering, reconstruction,
analysis.

A particular enphasys will be placed on
browsing data-stores down to object detail level
including support for 2D and 3D visualisation.

A final aim of such a coherent environment will be
the ability to move and share code across final analysis, reconstruction and triggers.

\section{CMS Software System}
  \vskip 8mm

\begin{figure}[htp]
  \centering
  \epsfig{file=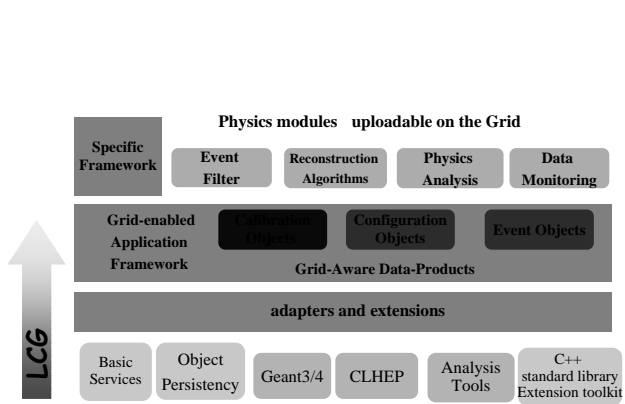,angle=-90,width=\linewidth} 
  \caption{CMS Software System and its compoments: the arrow shows
  the layers that LCG will enventually cover.} 
  \label{fig02}
\end{figure}

In order to achieve such a Coherent Analysis Environment
the software system has to provide a Consistent User interface on one
side and a coherent set of basic tools and mechanisms on the other.
CMS software system, presented in figure~\ref{fig02}, is based on a
modular and layered architecture\cite{cmsarch} centred around 
the COBRA\cite{COBRA} and IGUANA\cite{IGUANA} frameworks.
It relies on a high-quality service and utility toolkit provided by
the LCG project\cite{LCG} in the form of either certified and maintained external software components 
or as packages developed by the project itself.

The framework defines the top level abstractions, their behaviour
and collaboration patterns.
It comprises two components:
a set of classes that capture CMS specific concepts like
detector components and event features and a control policy
that orchestrates the instances of those classes 
taking care of the flow of control, module scheduling, input/output, etc.  
This control policy is tailored to the task in hand
and to the computing environment.

The physics and utility modules are written by detector groups and physicists.  
The modules can be plugged into the application framework at run time,
independently of the computing environment.  
One can easily choose between different versions of
various modules.  The physics modules do not communicate with each other
directly but only through the data access protocols that are part of the
framework itself.

Both the application framework and the service and utility toolkit
shield the physics software modules from the underlying technologies
which will be used for the computer services.
This will ensure a smooth transition to new technologies with changes
localised in the framework and in specific components of the service toolkit.

\section{Distributed Analysis}

\begin{figure}[htp]
  \centering
  \vskip 3mm
  \epsfig{file=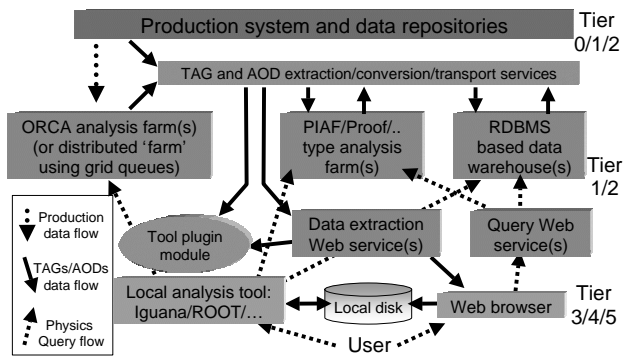,angle=-90,width=\linewidth}
  \caption{Distributed Analysis Architecture.} \label{fig03}
\end{figure}

Physicists analysing data will access a great variety of
sources and services ranging from simple histograms and n-tuples,
to Relational Data-Bases. Analysis activity may include simple ``tag''
selections, more complex SQL-like queries up to running CMS simulation,
calibration and/or reconstruction applications.
A combination of generic grid tools and specialised CMS tools will be
required to optimise the use of computing resources.
Figure~\ref{fig03} shows the current view of CMS architecture
for Distributed Analysis.
Physicists will use a single access point that will allow them to 
gain access to the required resources and services depending both on the 
task in hand and the nature of the data.
An Analysis Server, such as the Clarens Remote Dataserver
\cite{CLARENS} described in next section,
will act as mediator between the user, and his front-end application,
and the back-end services.  
These three-tier architecture realises a clean separation between
the user environment and the service environment.
This will enable the user to maintain a consistent
personalised environment while the various services
may be configured and deployed according to policies specific 
to each provider site.

\subsection{The Clarens Remote Analysis Server}

The Clarens Remote Dataserver \cite{CLARENS} project aims to build a wide-area
network client/server system for remote access to a variety of data and analysis
services.

The first service envisaged (and partly implemented) is analysis of
events from Spring 2002 production currently
stored in centralised Objectivity databases, but in future may be stored in
a combination of relational databases and/or files. Other services include remote
access to Globus functionality for non-Globus clients, including file transfer,
replica catalog access and job scheduling.

The concept of a ``server'' is currently the traditional interpretation
of the word, but might in future also refer to the multi-tiered data-GRID for
data access and analysis.

Communication between the client and server is conducted via the lightweight
XML-RPC \cite{XMLRPC} remote procedure call mechanism. This was chosen both
for its simplicity, good standardisation, and wide support by almost all programming
languages. Work is also underway to enable communication using the widely supported
SOAP \cite{SOAP} protocol.

The server is implemented using a modular combination of 
Python and C++  that executes directly
in the Apache \cite{APACHE} web server's address space. This increased performance
roughly by a factor of five w.r.t. a cgi-bin implementation. 
Clients connecting to the server are authenticated
using Grid certificates, while the server can likewise be authenticated by clients.
Client session data is tracked using the high-performance embedded Berkeley
database. Sessions can optionally be conducted over strongly encrypted SSL/TLS
connections. Authentication information is always strongly encrypted, irrespective
of whether the session is encrypted.

The modular nature of the server allows functionality to be added to a running
server without taking it off-line by way of drop-in components written in Python
of a combination of C++ and Python. An example module that allows browsing and
down-loading of histograms and Tag objects in an Objectivity database federation
has been implemented in this way.

The multi-process model of the underlying server (Apache) is uniquely suited
to handling large volumes of clients as well as long-running client requests.
These server processes are protected from other malicious or faulty requests
made by other clients as each process runs in its own address space.

There are currently three clients that are evolving along with the server: 

\itemsep 1pt \parskip 1pt 

\begin{itemize}
\item Python command line 
\item C++ command line client as an extension of the ROOT analysis environment 
\item Python GUI client in the SciGraphica analysis environment 
\end{itemize}
Support for the Java Analysis Studio client has been dropped in favour of the
ROOT client due to the latter's greater popularity in the CMS community. A web-based
front-end that supports the new authentication mechanism will be deployed in
the near future to replace the current non-authenticated version.

The command-line Python client may also be used directly in any Python-based
analysis environment such as the IGUANA scripting service.

\section{Data Challenges}

In order to test its Software system and its computing model
CMS is engaged in an aggressive program of ``data challenges'' of
increasing complexity:
\itemsep 1pt \parskip 1pt 
\begin{itemize}
\item Data Challenge `02 \\
Focus on High--Level Trigger studies
\item Data Challenge `04 \\
Focus on ``real-time'' mission critical tasks
\item Data Challenge `06 \\
Focus on distributed physics analysis
\end{itemize}
Even if each data challenge is focus on a given aspect, all encompass the whole data analysis process:
Simulation, reconstruction and statistical analysis running either as
organised production, end-user batch job or interactive work.

\subsection{Spring 2002 Production}

In Spring 2002 the CMS production team completed a production of Monte--Carlo
data~\cite{Prod02} for the DAQ TDR~\cite{DAQTDR}. 
Almost 6 million events were simulated
with the CMS GEANT3-based detector simulation program
CMSIM\cite{CMSIM} and
then digitised and reconstructed  under a variety of conditions with
the CMS Object Oriented reconstruction and
analysis program ORCA\cite{ORCA} based on the COBRA framework. 
About 20 regional centres participated
in this work running about 100 000 jobs for a total of 45 years CPU
(wall-clock) distributed over about 1000 cpus.
20TB of data were produced. Of these more than 10TB travelled on the
Wide area network.
More than 100 physics were involved in the final analysis.
PAW, and in a later stage ROOT, were the preferred analysis tools.
Less specialised tools, such as Mathematica and Excel, were also used for some
analyses.
IGUANA was used for interactive graphical inspection of detector
components and of single events.
The result of this work\cite{DAQTDR} was the
successful validation of CMS High--Level Trigger algorithms including
rejection factors, computing performance and functionality of the reconstruction-framework.

\subsection{Data Challenge in 2004 (DC04)}
The previous data challenge has concentrated on the simulation and digitisation
phase and the preparation of datasets for asynchronous analysis. 
With this challenge CMS intends to perform a large-scale test of the
computing and analysis models themselves. 
Thus we anticipate a pre-challenge period comprising the preparation
of the data samples, while the challenge itself consists of the reconstruction
and selection of the data at the T0 (Tier-0 computing centre at CERN), with
distribution to the distributed T1/T2 sites and synchronous analysis.

DC04~\cite{DC04} will run in the LCG~\cite{LCG} production grid. We anticipate that the
T0 part of the challenge is entirely CERN based, but the subsequent data replication
and analysis run in a common prototype grid environment. All T0-T2 resources
of the Data Challenge should be operating in a common GRID prototype\cite{claudio}.

In this data challenge, to be completed in April 2004, CMS will
reconstruct 50 million events in real time coping with a data rate equivalent to an 
event data acquisition running at 25 Hz for a luminosity
$2\times×10^{33} cms^{-2} s^{-1}$ for one month.
Besides this pure computation goal CMS will use this opportunity to
test the software system, the event model,
define and validate datasets for analysis,
identify reconstruction and analysis objects each physics group would like to
have for the full analysis,
develop selection algorithms necessary to obtain the required sample,
and prepare for ``mission critical'' analysis, calibration and alignment.

\section{Summary}
CMS is actively pursuing a development and test activity to be able to deploy
a fully distributed computing architecture in time for the LHC start-up. A distributed
production and analysis of 6 million Monte--Carlo events have been successfully
completed in 2002 involving 20 regional centres. Next major data challenge is scheduled
for the year 2004 and comprises a complexity scale equal to about 5\% of that
foreseen for LHC high luminosity running. It will run in the LCG production
grid and will be used to exercise a fully distributed physics analysis using
tools such as the Clarens Remote Dataserver running in conjunction with CMS
specific applications based on the COBRA and IGUANA frameworks.


\end{document}